\documentclass{article}

\usepackage{amsmath, amsfonts, amscd, amssymb}
\usepackage{amsthm}

\newcommand{\1}{\mathbf{1}}
\newcommand{\aaa}{\mathfrak{a}}
\newcommand{\aaaa}{\mathfrak{A}}

\newcommand{\avlst}{\mathfrak{S}}

\newcommand{\dff}{\sc}

\newcommand{\hhh}{\mathcal{H}}
\newcommand{\hhhh}{\mathcal{H}}

\newcommand{\lth}{N}

\newcommand{\mps}{\mathfrak{P}}

\newcommand{\ptt}{\sigma}
\newcommand{\pttr}{\Pi}
\newcommand{\pttt}{\Sigma}

\newcommand{\rrh}{\mathbf{\rho}}
\newcommand{\rsp}{\mathfrak{S}}

\newcommand{\yyyy}{\mathbf{S}}
\newcommand{\yyyyy}{\mathbf{A}}

\DeclareMathOperator{\loci}{Loc}

\title{A Combinatory-Algebraic Perspective\\ on
Multipartiteness, Entanglement\\ and Quantum Localization}

\author{Ioannis Raptis\thanks{\rm EU Marie Curie Fellow,
Theoretical Physics Group, Blackett Laboratory, Imperial College
of Science, Technology and Medicine, London SW7 2BZ, UK; e-mail:
i.raptis@ic.ac.uk} and Rom\`{a}n R. Zapatrin\thanks{\rm Quantum
Information Group, ISI, Villa Gualino, V.le S.Severo 65, 10133,
Torino, Italy; e-mail: zapatrin@isiosf.isi.it (address for
correspondence)}}

\date{}

\begin{document}

\maketitle

\pagestyle{myheadings}\markboth{\centerline {\small {\sc { Ioannis
Raptis and Rom\`an Zapatrin }}}}{\centerline {\footnotesize {\sc
{An Algebraic Perspective on Multipartiteness}}}}

\pagenumbering{arabic}

\begin{abstract}
We claim that both multipartiteness and localization of subsystems 
of compound quantum systems are of an essentially relative nature 
crucially depending on the set of operationalistically available 
states. In a more general setting, to capture the relativity and 
variability of our structures with respect to the observation 
means, sheaves of algebras may need be introduced. We provide the 
general formalism based on algebras which exhibits the relativity 
of multipartiteness and localization. 
\end{abstract}

\section{Prolegomena cum Physical Motivation}\label{sproleg}

The non-local behavior of quantum systems is virtually undisputed. 
There is ample experimental evidence suggesting that there exist 
quantum states of an essentially non-local nature, an issue which 
is verified by the statistics of observations. Entanglement is a 
crucial resource for quantum information processing and quantum 
communication. As it turns out, while it may be easy to produce 
non-entangled states, it is difficult to fabricate and maintain 
entangled ones. Recently, multipartite entanglement has been 
classified by the use of partitions of the set of subsystems 
\cite{durcirac}.

We may regard this as the first indication of the idea we wish to 
explore below, namely, {\em the relativity of the `property' for a 
subsystem to be observed---in effect, to be localized---somewhere}. 
Albeit, it is perhaps inadequate to just say relativity; one should 
also say {\em uncertainty} of some sort, as the position itself is 
created at the very moment of the preparation of the state. Quanta 
act not only non-locally, but also `{\em a-locally}' 
\cite{df2,rapzap1}, as if there is no given external physical 
space, fixed up-front as it were, to restrain their `quantum leaps 
of coherence and entanglement'. Even more iconoclastically, {\em 
space(time) is intuited to be `inherent' in quanta}, as it were, 
{\em created by them} \cite{rapzap2}. On the whole, it is more 
physical to maintain that space(time) and its mathematical analysis 
(topology and geometry) is a result of the algebraically 
represented (dynamical) relations between quanta rather than being 
fixed up-front, once and forever, by the theorist. We may distill 
all this to the following motto:

\vskip 0.07in

\centerline{{\em First comes the quantum, then space; not the
other way around}.}

\paragraph{Organization of the paper.} The paper is organized as follows.
We begin with an overview of how multipartiteness arises in
classical mechanics, what are the ways to recover it
operationally, and to what extent it is `absolute'. We point out
that in the case of the lack of availability of all states, it may
turn out that even classical systems may exhibit the `virtual'
character of their multipartite structure. This is just an
observation from classical statistics. Then we provide the
necessary basic definitions and recall how multipartiteness is
described in standard quantum mechanics. We observe that the
relativity of localization and entanglement already exists in the
usual quantum mechanics, so that there is no need to add to or to
remove from the standard theory essentially anything.

We then translate both classical and quantum multipartite issues
into a uniform algebraic language. This enables us to introduce
the notion of multipartite structures (MPS) on algebras in a way
that crucially depends on the set of available states. Then we
show how the structure of loci of subsystems---which we claim to
be the very structure of space(time)---{\em emerges} rather
naturally. At the end, we entertain the possibility that
observation-relativization and, concomitantly, locus-variability
may be mathematically modelled by (finitary) sheaves of (Rota, in
known cases \cite{rapzap1}) algebras over those loci-structures
(spectral topologies) inherent in quantum subsystems much in the
same way that has been accomplished for quantum spacetime foam
\cite{rapzap2} and gravity \cite{malrap1}.

\section{Classical compoundness}\label{sclassc}

In this section we describe two ways in which the compoundness of
classical systems may be treated as relative. We commence the
study of compoundness starting from classical systems. We show
that even at the classical level there are essentially two
different manifestations of the relativity of the notion of
multipartiteness. The first manifestation is due to
coarse-graining (grouping subsystems), while the second is related
to different decompositions of the available configuration space,
decompositions which depend on the scope of sets an experimenter
possesses at her disposal ({\it ie}, available or `experimentally
accessible' states). In this way we introduce the twofold
relativity of the notion of compoundness: compoundness based on
coarse-graining, and compoundness based on the choice of available
states, which we treat as being uncorrelated.

\paragraph{Cartesian product structures.} How can we actually
verify that a (classical) system is indeed multipartite? Let us
consider a simple model. On the one hand we have a classical
system whose configuration space $S$ consists of 9 points, while
on the other, two classical systems each having a 3-point
configuration space, say $M_1$ and $M_2$. The `first level'
mathematical description of them is identical: the configuration
space consists of 9 points, be it $S$ or $M_1\times{}M_2$. We are
not able yet to tell whether the first system is `here' and the
second one is `there' solely in terms of their configuration space
description. A Cartesian Product Structure (CPS) must be imposed
in order to draw such distinctions.

\begin{figure}[h!]
\begin{center} \begin{picture}(125,10)
\multiput(0,5)(7,0){9}{\circle{2}} \put(69,3.5){\mbox{$\mapsto$}}
\multiput(84,0)(5,0){3}{\circle{2}}
\multiput(84,5)(5,0){3}{\circle{2}}
\multiput(84,10)(5,0){3}{\circle{2}} \end{picture} \end{center}
\caption{An illustration of how CPS is imposed.} \label{fig1}
\end{figure}
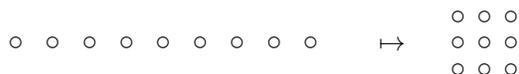

How can one impose a CPS on a set? A way to put it is by hand. In
our example this looks like making a rectangle from a line---see
Fig. \ref{fig1}.

\unitlength0.7mm
\newcounter{ccirc}
\setcounter{ccirc}{0}

\paragraph{Coarse-graining.} How many CPSs can one introduce on a
given finite set of cardinality $n$? An immediate answer is the
following: each possible CPS is associated with a particular
factorization $n=n_1\cdot{}n_2\cdots{}n_k$, where $k$ indicates
the number of subsystems associated with this particular CPS.
Then, up to permutations of the factors, all CPSs are in 1--1
correspondence with the factorizations of the number $n$.

Given a certain factorization $n=n_1\cdot{}n_2\cdots{}n_k$, we can
consider groups of factors as factors. In other words, we may not
`exhaust' the factorization, as it were, carry it to its `finest'
or `irreducible' level---{\it ie}, to $n$'s prime factors. Thus,
instead of an `{\em ultra-fine}' description of subsystems, we
consider coarser ones. This is the well known notion of
coarse-graining \cite{sorfinsub,rrzprg}

However, this approach is too rigid; in particular, it entails
that configuration spaces with a prime number of points have no
CPSs at all, and obviously the same holds in the quantum case, see
section \ref{scompqm}.

\paragraph{Compound systems with constraints.} Our first step
towards a realistic description of multipartiteness is to take
into consideration that the states and the observables of a
physical system may have different `accessibility status'. In
particular, some of them may turn out to be unavailable in our
experimental setting. When we have a configuration space equipped
with CPS, the states---which are probability distributions---may
be product or not.

Now, instead of declaring up-front a CPS on a set, let us try to
go the other way around and consider the case when two classical
parties are far away from each other so that it takes considerable
effort to make their states correlated. That means, we distinguish
between states which are `easy' for us to prepare and those which
are not. The formulation of the inverse problem beckons: given a
configuration space $S$ and a collection of states, how can one
recover a CPS $S=M_1\times{}M_2$ such that the available states
are product or weakly correlated ones?

This is a purely statistical problem which can be solved by using
principal component analysis. The idea is that the distributions
are viewed as points in a multi-dimensional space and then an
eigenvalue problem is solved, so that the number of resulting
vectors with sufficiently large eigenvalues gives the number of
principal factors, which in our case may be viewed as the number
of subsystems. What is swept out by cutting `sufficiently small'
eigenvalues are correlated states which, according to our setting
of the problem, are rare and pragmatically inaccessible.

Suppose that we have a `really' bipartite classical system, but
one which also has constraints. This means that there are states
in the Cartesian product configuration space which are not
accessible for us. Therefore, the `effective configuration space'
will no longer carry a Cartesian product structure; rather, it
will be its proper subset. 

Our claim is the following. CPSs can be introduced irrespectively
of the extent to which we can decompose the number of points of
the available configuration space. The only relevant issue is the
analysis of correlations. Therefore, from now on we change the
spelling of `C' in CPS from `Cartesian' to `classical' and CPSs
become {\dff classical product structures}.

\paragraph{Constrained systems viewed algebraically.} Now, let us
base our considerations on an algebraic ground. Suppose we have a
bipartite system, and we measure local observables $A_1$ and
$A_2$. If the state $\rrh$ in which we measure them is a product
one, the mean value of the product equals the product of the
values:

\begin{equation}\label{eloccom}
\rrh(A_1\cdot{}A_2) = \rrh(A_1)\cdot{}\rrh(A_2)
\end{equation}

This may be regarded as a characteristic property.  If we have two
algebras $\aaa_1$ and $\aaa_2$ of local observables, then for any
product state and any $A_1\in\aaa_1$, $A_2\in\aaa_2$, condition
\eqref{eloccom} holds.

Our idea is to forget for the time being about the CPS on the
configuration space and start from a given set $P$ of states which
we {\em declare} to be product. Then we may take two subalgebras
$\aaa_1$ and $\aaa_2$, and ask whether for any $\rrh\in{}P$
\eqref{eloccom} holds. If the answer is yes, then operationally,
from the point of view of available observations, we are dealing
with a bipartite system. Note that this approach is perfectly
applicable to systems with constraints; when the effective
configuration space has no product structure, the multiplicativity
\eqref{eloccom} still holds!

Note that when we have a CPS on a set, we can consider local
algebras of observables. In turn, each local algebra has its set
of points. What is the relation between the global algebra of
observables and local ones, between the overall configuration
space and the configuration spaces of the individual systems? The
answer is known. The global algebra is (in general, a superset of)
the tensor product of local algebras, and the overall
configuration space is (in general, a superset of) the Cartesian
product of individual configuration spaces.

This gives us a clue to introduce multipartiteness in an algebraic
fashion. Namely, take a collection of subalgebras of the overall
algebra of observables. For any state $\rrh$ we may ask if the
analog of \eqref{eloccom} holds. Note that this is not an attempt
to treat states as points, which is meaningless, since for two
local observables \eqref{eloccom} does not hold in general. That
is why we introduce the term {\em loci} instead of points.

\paragraph{Classical product structures.} In classical mechanics
the algebra of observables is a (commutative) algebra of smooth
functions defined on the configuration space. Due to Gel'fand
duality we may generalize the ideas of section \ref{sclassc} and
introduce CPS in a purely algebraic way, that is, with no {\it a
priori} reference to the underlying geometrical configuration
space (manifold), which only later will be recovered by the
representation theory of the algebras employed much in the same
way we did for spacetime foam in \cite{rapzap2}. A {\dff Classical
Product Structure} (CPS) $\mps$ is a set of unital subalgebras.

\[
\mps=\left\{\aaa_1,\ldots,\aaa_n\right\}
\]

The algebras $\aaa_i$ forming this set are said  to be {\dff
algebras of local observables}. The labels which mark each
subalgebra is called {\dff locus}. This definition looks at first
sight counterintuitive as it apparently disagrees with the
standard viewpoint: {\bfseries (i)} we do not require the local 
subalgebras to intersect only at the unit element of the embracing 
algebra and the reason for this is because they may be 
indistinguishable from the unit element when we have a limited 
number of states at our disposal. {\bfseries (ii)} we do not 
require the local subalgebras to comprise the whole algebra of 
observables, this too reflecting our `{\em local experimental 
ignorance}' concerning the totality of properties of the quantum 
system that can in principle be observed.

\paragraph{State-CPS duality.} Given an algebra $\aaaa$, a CPS
$\mps=\{\aaa_1,\ldots,\aaa_n\}$ and a state $\rrh$ on it, we say
that $\rrh$ is {{\dff product with respect to} $\mps$ whenever the
generalization of the condition \eqref{eloccom} holds:

\begin{equation}\label{edefacps}
\forall i\quad \forall A_i\in\aaa_i \quad
\rrh\left(\prod_i{}A_i\right) = \prod_i{}\rrh\left(A_i\right)
\end{equation}

\noindent Therefore we may encounter the following situation.
Suppose we have a set of states, which we may regard as being
`easily available'. Then, it may happen that there are several
inequivalent CPSs with respect to which these states are product.

Our claim is the following. Even in the classical case, in a 
situation where we have a restricted set of states at our disposal, 
all product structures can be treated as full fledged 
multipartite(ness) as we have no operational means to single out, 
`prefer', or discriminate between particular states. Now we may 
approach the basic claims of our paper. Even in a classical 
situation the following hold. {\em What creates observable 
multipartiteness?} The set of available states. {\em Where is the 
multipartite structure genuinely imposed?} On the algebra of 
observables.

\section{Compoundness in quantum mechanics}\label{scompqm}

In this section we show that all the issues concerning the
relativity of classical multipartiteness are still effective in
the quantum case. Furthermore, the variety of quantum multipartite
structures acquires a new, continuous degree of freedom. The main
difference between the classical and the quantum case is that the
condition for a state to be product is replaced by separability.

\paragraph{The relativity of multipartite entanglement.} Begin with 
basic definitions. Given a state of a composite $\lth$-partite 
system $\yyyy$, denote its density matrix by $\rrh$. A density 
matrix $\rrh$ is called {\dff product} if it can be represented as 
a tensor product of density matrices of subsystems $\rrh = \rrh_1 
\otimes\ldots\otimes \rrh_\lth$. A state $\rrh$ is {\dff separable} 
if its density matrix is a convex or `incoherent' ({\it ie}, a 
classical probabilistic) linear combination of product ones.

\begin{equation}\label{e03s}
\rrh = \sum p_\alpha \rrh_1^\alpha \otimes\ldots\otimes
\rrh_\lth^\alpha
\end{equation}

\noindent with $p_\alpha\ge0$ and $\sum{}p_\alpha=1$.

\medskip

The states which are not separable are called {\dff entangled}.

The first hint we wish to give towards introducing our algebraic 
picture of compoundness is based on the following rationale: when 
we have several parties in an entangled state we must consider them 
as a single party---an inseparable entity. At the same time, what 
right have we got to still call this `coherent whole', `several 
entangled parties'? Presumably, because in principle we also have 
at our disposal other states, which are separable and, {\it a 
fortiori}, which can separate or individuate the constituent 
parties.

\paragraph{Coarse-graining.} To introduce the quantum version of
coarse-graining in CPS we proceed in close analogy with the
classical case, the only, albeit essential, difference being that
the aforementioned separability condition should also be taken
into account.

So, let us weaken the condition for states of a composite system
$\yyyy$ to be product and separable. Namely, instead of requiring
it to be  product with respect to $\hhhh = \hhh_1
\otimes\cdots\otimes \hhh_\lth$, we make this condition {\em
relative} to a partition $\pttt$ of the set $\yyyyy$ of subsystems
of $\yyyy$, that is, with respect to a particular decomposition of
the set of subsystems. In a sense, we are allowed to relax in this
way the product and separable states of a composite system,
because, as explained earlier, they are precisely the `least
quantum' ones ({\it ie}, non-entangled).

Given a partition $\pttt=\{\ptt_1,\ldots,\ptt_M\}$ of $\yyyyy$ and
a density matrix $\rrh$ in the state space of $\yyyy$, $\rrh$ is
called {\dff $\pttt$-product} whenever it can be represented as a
tensor product $\rrh = \rrh_{\ptt_1} \otimes\cdots\otimes
\rrh_{\ptt_M}$ and {\dff $\pttt$-separable} if it is a convex
combination of $\pttt$-product states:

\begin{equation}\label{e04}
\rrh = \sum p_\alpha \rrh_{\ptt_1}^\alpha \otimes\ldots\otimes
\rrh_{\ptt_\lth}^\alpha
\end{equation}

\noindent In other words, \eqref{e04} means that we can prepare
$\rrh$ as an ensemble of mixed states located at sites
$\ptt_1,\ldots,\ptt_M$.

Given a state $\rrh$, we may now ask for each partition $\pttt$ of
the set $\yyyyy$ of subsystems of $\yyyy$ whether $\rrh$ is
$\pttt$-separable or not. As a result, we obtain the set
$\pttr(\rrh)$ of partitions of $\yyyyy$ with respect to which
$\rrh$ is separable \eqref{e04}:

\begin{equation}\label{e05}
\pttt\in\pttr(\rrh) \quad\Leftrightarrow\quad \rrh \quad \mbox{is
$\pttt$-separable}
\end{equation}

\noindent The set of all partitions of a given set has a natural
ordering `$\preceq$', which represents acts of coarse-graining
those partitions. In order to specify all partitions with respect 
to which a given state $\rrh$ is separable, we only have to find 
{\em maximal} ones with respect to `$\preceq$'. This may serve a 
base for geometrico-algebraic invariants for multipartite 
entanglement, for details of which the reader may refer to 
\cite{rrzoviedo}.

\paragraph{Tensor product structures (TPSs).} In Section
\ref{sclassc} we introduced the variety of CPS for classical
systems. This admits an immediate generalization to quantum
systems which was carried out by Svozil \cite{karl02} and Zanardi
\cite{paolo}. Let us briefly review it. First, following
\cite{karl02}, take an arbitrary basis in a Hilbert space $\hhh$
of dimension $n$ with {\em no a priori given} tensor product
structure. Take any factorization $n=n_1\cdot\cdots\cdot{}n_k$ and
associate with it $k$ partitions of the set of basis vectors. The
partitions must be such that each basis vector could be
represented as an intersection of appropriate elements of each
partition. In other words, these partitions should form
independent Boolean algebras. Then, taking an element of a
partition, we may view it as a qubit (in generalized sense with an
arbitrary number of values). To get the next degree of freedom in
producing different TPSs, apply according to \cite{paolo} a global
unitary transformation $U:\hhh\to\hhh$, which will yield us
isomorphic, but different (with respect to, say, observables),
qubit structures. In the sequel, the subsystems of the associated
multi-party decomposition will be referred to as {\em virtual}.

So, by now we have completed the description of both classical and
quantum  multipartite systems. What we have done was essentially
to show how {\em given a compound system} we can describe it in
different ways and in some sense `modify' its compoundness. The
main goal of our paper is, however, to provide the appropriate
algebraic machinery for {\em creating} compoundness. We address
this issue in the next section.

\section{Compoundness from an algebraic
perspective}\label{salgcompd}

As it has been pointed out above, both classical and quantum
compound systems exhibit some kind of relativity of their
multipartite structure. In this section we take a step further and
turn the construction the other way around. Starting from a given
set of observables and states---in fact, this is perhaps the only
way that our approach may qualify as being operationalistic
proper---we {\em create} rather than (re)construct (and this is
our main claim here!) the full fledged multipartite structure of
compound quantum systems. It is full fledged, because no
experiment can be devised to discriminate between our `fake'
multipartiteness and the purported `real' one. As a matter of
fact, we abide to the stronger statement that there is simply no
`real' multipartiteness at all.

We shall describe both classical and quantum systems by using
algebraic means. That is, we shall regard {\em `algebras of
observables'} as primary theoretical objects. As a result, the
geometrical configuration and state spaces will turn out to be
just {\em representation spaces} for those algebras. This is in
line with a generalized notion of {\em Gel'fand duality}
\cite{rapzap2}. Note that the term `algebra of observables' is
rather broadly and heuristically used here, as only its
self-adjoint elements correspond to observables proper. The
algebra itself is broader as it embodies both observables and {\em
evolutions}. Recall the basic definitions.

Now we are going to provide a quantum, noncommutative analog of
the classical product structures. The language of algebras is so
adequate for it that we do not have to introduce practically
anything new. The main difference between the classical and the
quantum case is that the variety of MPS in the latter becomes much
broader and richer than in its classical counterpart.

\paragraph{Virtual multipartite structures.} A collection $\mps$

\begin{equation}\label{edefmps}
\mps \;=\; \{ \aaa_i \}_{i\in \loci(\mps)}
\end{equation}

\noindent of unital subalgebras of $\aaaa$ which are closed with
respect to taking double commutant $\forall i \:
(\aaa_i)^{cc}=\aaa_i$ is called {\dff virtual multipartite
structure} (MPS).

There is a partial order on MPS which enables us to represent the
possibility of coarse-graining in both classical and quantum MPSs.
It is introduced by analogy with that on partitions.  Namely, we
say that an MPSs $\mps=\{ \aaa_i \}_{i\in \loci(\mps)}$ is {\dff
coarser} than $\mps'=\{\aaa_i\}_{i\in\loci(\mps)}$ (denote it by
$\mps\preceq\mps'$) if we can partition $\mps'$ so that the span
of each element of the partition is a subalgebra of appropriate
$\aaa_i$ from $\mps$.

Furthermore, this is a lattice ordering. In fact, if we have a set
of MPSs $\mps^1,\ldots,\mps^r$ we may take all possible
intersections of all subalgebras from all MPPSs. The result will
be again an MPS which will be the least upper bound of
$\mps^1,\ldots,\mps^r$ with respect to the relation "$\preceq$"

\[
\mps^1\vee\ldots\vee\mps^r \:=\:
\bigcap_{i=1}^r\bigcap_{k^i\in\loci(\mps^i)}\left\{\aaa^i_{k^i}\right\}
\]

\noindent The lattice structure can then be used to recover the
loci.

\paragraph{The state-MPS duality.} We emphasize that we {\em did
not} require the elements of different `local' subalgebras  to
commute. This for instance is in striking contrast to the usual
(involutive) observable algebras of (relativistic) quantum matter
systems which are already localized in Minkowski space, as the
geometry of the fixed background spacetime dictates the
`commutativity {\it vis-\`a-vis} local causality' properties of
the corresponding algebras (Einstein Locality). Here, exactly
because we do not posit up-front an ambient base localization
space(time), we are not {\it a priori} constrained by Einstein
Locality and, as a result, quantum non-locality effects do not
surprise us. This is the crucial point of our approach---{\em we
shall require commutativity only on certain states which we treat
as being `available'}. Operationally, that means that the values
of all `local' variables should be independent random (stochastic)
variables, but under the proviso that {\em the system is in an
available state only}. We thus define the relation $\rsp$ of
separability between a state $\rrh$ and an MPS $\mps$ as follows

\begin{equation}\label{edefrelsep}
\rrh\,\rsp\,\mps \quad\Leftrightarrow\quad \forall a_i\in\aaa_i \:
\forall a_j\in\aaa_j \:\: \rrh(\aaa_i\cdot\aaa_j)=
\rrh(\aaa_i)\cdot\,\rrh(\aaa_j)
\end{equation}

\noindent Note that the rhs of \eqref{edefrelsep} is well defined
as both $\aaa_i,\aaa_j$ belong to the total algebra $\aaaa$. The
analog of local observables in standard tensor stutures are the
operators of the form
$\1\otimes\cdots\otimes{}A_i\otimes\cdots\otimes\1$. Then, by
analogy with \eqref{e05} we introduce the duality using  the same
notation $\pttr(\rrh)$ for the appropriate set of MPSs:
\begin{equation}\label{e05a}
\mps\in\pttr(\rrh) \quad\Leftrightarrow\quad \rrh \quad \mbox{is
$\mps$-separable}
\end{equation}

\paragraph{Recovering the loci.} As claimed above, we start with an algebra  $\aaaa$ of observables
and the set $\avlst$ of available states. With any state
$\rrh\in\avlst$ we can associate the set $\pttr(\rrh)$ of MPSs
with respect to which $\rrh$ is separable.

The loci of any of MPS from $\pttr(\rrh)$ are still thought of as
groups of elementary subsets. Now, using the lattice structure on
the set of all MPSs, we can form the supremum

\begin{equation}\label{erecov}
\mps_\avlst \:=\:
\bigvee_{\rrh\in\avlst}\bigvee_{\mps\in\pttr(\rrh)}\mps
\end{equation}

\noindent which is still an MPS. This is exactly what provides us
with `points'---namely, the loci $\loci(\mps_avlst)$  are treated
as ultimately indivisible with respect to the given set of
accessible states, which in turn one may identify with the
available (microscopic) energies.

How and why does relativity come about? When we broaden the range
of available states (as it were, increase the energy of
microscopic resolution) the number of terms in \eqref{erecov} may
only increase, therefore the MPPSs $\mps_\avlst$ may become finer.
This in turn means that its loci `decay' or break down to
`smaller' ones. The benefit we get from this construction from the
point of view of quantum computing is that when the loci are
defined (created) we can then directly apply Svozil's
\cite{karl02} partition scheme in order to reconstruct qubits.
These are {\em fully fledged qubits} viewed purely operationally.

We conclude the paper by noting briefly that in the already worked
out case where the `compound' system is {\em quantum spacetime},
the aforesaid algebras have been seen to be (non-involutive) {\em
Rota algebras}, while the set of loci was endowed with a so-called
{\em spectral Rota topology} by employing a variant of the idea of
{\em Gel'fand duality} coined {\em Gel'fand spatialization}
\cite{rapzap1,rapzap2}. This theoretical scenario is supposed to
represent a combinatory-algebraic description of (the kinematics
of) {\em spacetime foam}---the conception of the spacetime
microtopology as being a quantum observable \cite{rapzap2}. At the
same time, it has been intuited that, by further regarding these
algebras as noncommutative {\em local rings}, their
sheaf-theoretic localizations over their Gel'fand spectra can
capture, by entirely algebraico-categorical means, notions of
relativity and dynamical variability of quantum discretized
spacetime \cite{malrap1}.

Thus, similarly here we intuit that the structure of loci may be
endowed with a suitable spectral topology and, concomitantly,
their relativity and dynamical variability can be captured by
considering sheaves of the relevant associative algebras over
these topological spaces. Such raw analogies may provide the
fertile ground for future investigations in which ideas from
quantum computation proper can be brought closer to ones from
quantum spacetime and gravity.


\begin{thebibliography}{99}

\bibitem{durcirac}
D\"ur, W., and Cirac, J.-I., Physical Review A, {\bf 61,} 042314
(2000)

\bibitem{df2} Finkelstein, D. R., {\itshape Quantum Relativity: A Synthesis of the Ideas of Einstein
and Heisenberg}, Springer-Verlag, Berlin-Heidelberg-New York
(1996)

\bibitem{malrap1} Mallios, A. and Raptis, I.,
{\itshape Finitary Spacetime Sheaves of Quantum Causal Sets:
Curving Quantum Causality}, International Journal of Theoretical
Physics, {\bf 40}, 1885 (2001); gr-qc/0102097

\bibitem{rapzap1} Raptis, I. and Zapatrin, R. R.,
{\itshape Quantization of discretized spacetimes and the
correspondence principle}, Int. J. Theor. Phys., {\bf 39}, 1
(2000); gr-qc/9904079

\bibitem{rapzap2} Raptis, I. and Zapatrin, R. R.,
{\itshape Algebraic description of spacetime foam}, Classical and
Quantum Gravity, {\bf 20}, 4187 (2001); gr-qc/0102048

\bibitem{sorfinsub}
Sorkin,  R., {\itshape Finitary Substitute for Continuous
Topology,} International Journal of Theoretical Physics, {\bf 30},
923 (1991)

\bibitem{karl02} Svozil, K.,
{\itshape Quantum information in base $n$ defined by state
partitions}, Physical Review Letters, {\bf A666}, 044306 (2002);
quant-ph/0105031

\bibitem{paolo}
Zanardi, P., {\itshape Virtual Quantum Subsystems }, Physical
Review Letters, {\bf 87}, 077901 (2001); quant-ph/0103030

\bibitem{rrzprg} Zapatrin, R. R., {\itshape Pre-Regge Calculus: Topology Via Logic},
International  Journal  of  Theoretical Physics, {\bf 32}, 779,
(1993)

\bibitem{rrzoviedo} Zapatrin, R. R.,
{\itshape Combinatorial Topology Of Multipartite Entangled
States,}, Journal of Modern Optics (2002), to appear; eprint
quant-ph/0207058

\end{thebibliography}
\end{document}